\documentclass[12pt]{article}
\begin{document}

\begin{center}
{\bf Internal Symmetry Group and Density Matrix of Fields with Spins 0, 1 } \\
\vspace{5mm} S.I.Kruglov
\footnote{E-mail: skruglov23@hotmail.com} \\
\vspace{3mm}
\textit{International Educational Centre, 2727 Steeles Ave.West, Suite 202,
\\Toronto, Ontario, Canada M3J 3G9 }\\
\vspace{5mm}
\end{center}

\begin{abstract}
The internal symmetry group U(3,1) of  the neutral vector fields with two spins
0 and 1 is investigated. Massless fields correspond to the generalized Maxwell
equations with the gradient term. The symmetry transformations in the
coordinate space are integro-differential transformations. Using the method
of the Hamiltonian formalism the conservation tensors are found, and the quantized
theory is studied. The necessity to introduce an indefinite metric is shown.
The internal symmetry group $U(3,1)$ being considered, after the transition to
electrodynamics, reduces to the $U(2)$ group. It is shown that the group of
dual transformations is the subgroup of the group under consideration.
 All the linearly independent solutions of the equation for a free particle
 obtained in terms of the projection matrix-dyads.

\end{abstract}

\section{Canonical formalism}

In the general case, without any constraints, the vector field $B_{\mu}(x)$
realizes the $\left( 0,0\right) \oplus \left( 1/2,1/2\right) $ representation
of the Lorentz group and describes four degrees of freedom which correspond to
states with spins $s=0$ and $s=1$ (with three spin projections $s_z=0,$ $\pm1$).
The massive field functions $B_{\mu}(x)$ satisfy the Klein-Gordon-Fock equation
\begin{equation}
\left( \partial _\mu ^2-m^2\right) B_{\nu}(x) =0,  \label{1}
\end{equation}

where $\partial_\mu =(\partial /\partial x_m, \partial /i\partial t$,
$B_{\nu}(x)=(B_{n}(x),iB_{0}(x))$.

The corresponding Lagrangian for the neutral fields can be rewritten
as follows (within unimportant divergent-type terms):
\begin{equation}
{\cal L}=-\frac 12\left[ \left( \partial _\mu B_\nu \right) ^2+m^2B_\mu
^2\right].  \label{2}
\end{equation}

The Lagrangian (2) can be connected also with the Stueckelberg formulation
of the vector field [1]. A Lagrangian of the form (2) also was used
[2] in a gauge-invariant formulation for a massive neutral vector field.

Eq.(1) can be represented in the form of first-order equations [3]
\[
\partial _\nu \psi _{\mu \nu}-\partial _\mu \psi _0+m^2\psi _\mu =0,
\]
\[
\psi _{\mu \nu}=\partial _\mu \psi _\nu-\partial _\nu \psi _\mu ,
\]
\begin{equation}
\psi _0=\partial _\mu \psi _\mu ,  \label{3}
\end{equation}

with $\psi _\mu =B_\mu $, $\psi _0 =\partial_\mu B_\mu $. In the case
$m=0$ equations (3) are the generalized Maxwell equations with the
gradient term (see [4-6]).
Now we investigate the symmetry group of the four-component neutral vector
field $B_\mu $ which describes two spins $0$, $1$ (without the Lorentz
condition, i.e. $\partial _\mu B_\mu \neq 0$ (see Eq.(1)) with the
Lagrangian (2) [7]. We call this field ``the Stueckelberg field'' [8].

Let us consider the transformations $\Lambda =(\Lambda _{\mu \nu })$ of the
field functions $B_\mu ^{\prime }(x)=\Lambda _{\mu \nu }B_\nu (x),$ which
belong to the group $SO(3,1)$ but which leave the coordinates $x_\mu $
unchanged, i.e. $x_\mu ^{\prime }=x_\mu.$ This is different case from that
of the Lorentz group where the coordinates $x_\mu $ are transformed. It is
easy to verify that the Lagrangian (2) is invariant under this group of
symmetry transformations as $\Lambda _{\mu \alpha }\Lambda _{\mu \beta
}=\delta _{\alpha \beta }$ and $B_\mu ^{\prime 2}(x)=B_\mu ^2(x)$. In
accordance with the Noether theorem [9,10] the invariance of the action
integral under the group of the transformations under consideration one
yields the law of conservation of the following antisymmetric tensor
\begin{equation}
S_{\mu [\alpha \beta ]}=B_\beta \partial _\mu B_\alpha -B_\alpha \partial
_\mu B_\beta.  \label{4}
\end{equation}

This tensor coincides with the density of spin momentum corresponding to
the conservation law of spin momentum (4): $\partial _\mu S_{\mu [\alpha
\beta ]}=0$.

The invariance of the action integral (2) under the Lorentz
transformations of the coordinates $x_\mu ^{\prime }=L_{\mu \nu }x_\nu $
induces transformations of the field functions $B_\mu ^{\prime }(x^{\prime
})=L_{\mu \nu }B_\nu (x)$ which lead to a conservation law for angular
momentum $M_{\mu [\alpha \beta ]}$ [9,10], this being the sum of the
orbital, $T_{\mu [\alpha }x_{\beta ]}$, and spin, $S_{\mu [\alpha \beta ]}$,
momenta:
\[
T_{\mu \nu }=-\left( \partial _\mu B_\alpha \right) \left( \partial _\nu
B_\alpha \right) -{\cal L}\delta _{\mu \nu },
\]
\begin{equation}
M_{\mu [\alpha \beta ]}=T_{\mu [\alpha }x_{\beta ]}+S_{\mu [\alpha \beta ]}.
\label{5}
\end{equation}

Therefore we also have here the law of conservation of orbital angular
momentum: $\partial _\mu T_{\mu [\alpha }x_{\beta ]}=0$. It should be noted
that the Lorentz calibration $\partial _\mu B_\mu =0$ is not invariance
under our group of symmetry $(\Lambda _{\mu \nu })$. As the Lorentz
calibration extracts the pure spin $1$ of particles we come to the
conclusion that the laws of conservation of the orbital, $T_{\mu [\alpha
}x_{\beta ]}$, and spin, $S_{\mu [\alpha \beta ]}$, angular momenta
separately: $\partial _\mu T_{\mu [\alpha }x_{\beta ]}=0$, $\partial _\mu
S_{\mu [\alpha \beta ]}=0$ are due to the presence of two spins $0$ and $1$;
i.e. multi-spin $0,1$.

Now we consider a wider group of symmetry using the method of the
Hamiltonian formalism.

The generalized coordinates in this scheme are
\begin{equation}
q_\mu (x)=B_\mu (x). \label{6}
\end{equation}

The density of the momenta found from Eq.(2) is given by
\begin{equation}
\pi _\mu (x)=\frac{\partial {\cal L}(x)}{\partial \stackrel{\cdot }{q}
_\mu (x)}=\stackrel{\cdot }{q}_\mu (x),  \label{7}
\end{equation}

where $\stackrel{\cdot }{q}_\mu (x)=\partial q_\mu (x)/\partial t.$ The
density of the Hamiltonian is defined by the relationship
\begin{equation}
{\cal H}(x)=\pi _\mu (x)\stackrel{\cdot }{q}_\mu (x)
-{\cal L}(x).  \label{8}
\end{equation}

Inserting the density of the Lagrangian (2) and the densities of the
momenta (7) into Eq.(8) we arrive at
\begin{equation}
{\cal H}(x)=\frac 12\biggl [\left( \partial _mB_\mu (x)\right) ^2+
\stackrel{\cdot }{B}_\mu ^2(x)+m^2B_\mu ^2(x) \biggr ].  \label{9}
\end{equation}

In momentum space the real fields $B_\mu (x)$ are given by
\begin{equation}
B_\mu (x)=L^{-3/2} \sum_{{\bf k}} \left[ B_\mu ({\bf k}
,t)e^{i{\bf kx}}+B_\mu ^{+}({\bf k},t)e^{-i{\bf kx}}\right] ,
\label{10}
\end{equation}

where $L$ is the normalizing length so that the energy of a quantum is $
k_0=2\pi /L$ and the normalizing volume is $V=L^3$; $B_\mu ^{+}({\bf k},t)
$ is the Hermitian conjugated quantity. The time dependence of fields in
the momentum space is
\begin{equation}
B_\mu ({\bf k},t)\sim e^{-ik_0t},\hspace{0.5in}B_\mu
^{+}({\bf k},t)\sim e^{ik_0t},  \label{11}
\end{equation}

where $k_0^2={\bf k}^2+m^2.$ Taking into account Eqs.(9), (10) the
Hamiltonian $H=\int {\cal H}(x)d^3x$ takes the form
\begin{equation}
H=\sum_{{\bf k}}2k_0^2B_\mu ({\bf k},t)B_\mu ^{+}({\bf k},t). \label{12}
\end{equation}

Introducing the canonical variables in the momentum space
\[
q_\mu ({\bf k},t)=B_\mu ({\bf k},t)+B_\mu ^{+}({\bf k},t),
\]
\begin{equation}
\pi _\mu ({\bf k},t)=\stackrel{\cdot }{q}_\mu({\bf k}
,t)=-ik_0\left[ B_\mu ({\bf k},t)-B_\mu ^{+}({\bf k},t)\right] ,
\label{13}
\end{equation}
the Hamiltonian (12) is rewritten as
\begin{equation}
H=\frac 12 \sum_{{\bf k}}\left[ \left( \pi _\mu ({\bf k}
,t)\right) ^2+k_0^2\left( q_\mu({\bf k},t)\right) ^2\right] =\sum_{
{\bf k}}2k_0^2\varphi ^{+}\varphi,  \label{14}
\end{equation}

\begin{equation}
\varphi =\left(
\begin{array}{c}
B_1({\bf k},t) \\
B_2({\bf k},t) \\
B_3({\bf k},t) \\
iB_0({\bf k},t)
\end{array}
\right) ,\hspace{0.3in}\varphi ^{+}=\left(
\begin{array}{cccc}
B_1^{+}({\bf k},t) & B_2^{+}({\bf k},t) & B_3^{+}({\bf k},t) &
iB_0^{+}({\bf k},t)
\end{array}
\right) .  \label{15}
\end{equation}

It is obvious that the Hamiltonian (14) is invariant (see [11, 12]])
under the transformations of the pseudounitary group $U(3,1)$ (for real
fields $B_m({\bf k},t)$, $B_0({\bf k},t)$):
\begin{equation}
\varphi ^{\prime }=U\varphi ,\hspace{0.5in}\varphi ^{+^{\prime }}=\varphi
^{+}U^{+},  \label{16}
\end{equation}

where the complex $4\times 4$-matrix $U$ obeys the equation $U^{+}U=1$. The
matrix of the infinitesimal transformations (16) can be represented as
\begin{equation}
U=\left( 1-i\omega _0\right) I_4,+\omega _{[\mu \nu ]}I_{[\mu \nu ]}-\omega
_{(\mu \nu )}I_{(\mu \nu )},  \label{17}
\end{equation}

where the antisymmetric $I_{[\mu \nu ]}$ and symmetric $I_{(\mu \nu )}$
generators of the group are given by
\begin{equation}
I_{[\mu \nu ]}=\varepsilon ^{\mu ,\nu }-\varepsilon ^{\nu ,\mu },
\hspace{0.5in}I_{(\mu \nu )}=i\left( \varepsilon ^{\mu ,\nu }+\varepsilon
^{\nu ,\mu }-\frac 12\delta _{\mu \nu }\right) .  \label{18}
\end{equation}

Here $I_4$ is the unit $4\times 4$-matrix and the elements of the entire
algebra $\varepsilon ^{\mu ,\nu }$ satisfy the relations [9]
\begin{equation}
\left( \varepsilon ^{\mu,\nu}\right) _{\alpha\beta}=\delta _{\mu\alpha}
\delta _{\nu\beta},
\hspace{0.5in}\varepsilon ^{\mu,\nu}\varepsilon ^{\alpha,\beta}=
\delta _{\nu\alpha}\varepsilon^{\mu,\beta},  \label{19}
\end{equation}
where indexes $\mu,\nu,\alpha,\beta=1,2,3,4$.
The parameters of the $U(3.1)$ group obey the equations $\omega _0^{*}
=\omega _0$, $\omega _{[ab]}^{*}=\omega _{[ab]}$, $\omega _{[a4]}^{*}
=-\omega _{[a4]}$, $\omega _{(ab)}^{*}=\omega _{(ab)}$,
$\omega _{(a4)}^{*}=-\omega _{(a4)}$ ($a,b=1,2,3$). The generators of
the $U(3.1)$ group: $iI_4$, $I_{[\mu \nu ]}$,
$I_{(\mu \nu )}$ have the correct commutation relations [13] as may be
verified using Eqs.(19). The $I_{[\mu \nu ]}$ are the generators of the
subalgebra corresponding to the $SO(3,1)$ group and $iI_4$ corresponds to
the subgroup $U(1)$. Transformations (16) act in the space of the
functions (15) and the momentum ${\bf k}$ is not transformed.

To get the integral of motion (corresponding to the group of symmetry
(16) in the framework of this formalism one needs to consider the
canonical transformations which leave the Hamiltonian (14) invariant.
Taking into account Eqs.(17), (18), infinitesimal transformations (16) can
be cast into the form
\[
\delta B_\mu ({\bf k},t)=-i\omega _0B_\mu ({\bf k},t)+2\omega _{[\mu
\nu ]}B_\nu ({\bf k},t)-
\]
\begin{equation}
-2i\left[ \omega _{(\mu \nu )}-\frac 14\omega _{(\alpha \alpha )}\delta
_{\mu \nu }\right] B_\nu ({\bf k},t),  \label{20}
\end{equation}
\[
\delta B_\mu ^{+}({\bf k},t)=i\omega _0B_\mu ^{+}({\bf k},t)+2\omega
_{[\mu \nu ]}B_\nu ^{+}({\bf k},t)+
\]
\begin{equation}
+2i\left[ \omega _{(\mu \nu )}-\frac 14\omega _{(\alpha \alpha )}\delta
_{\mu \nu }\right] B_\nu ^{+}({\bf k},t).  \label{21}
\end{equation}

With the help of Eq.(13), transformations (20), (21) can be rewritten
in the canonical form
\[
\delta q_\mu({\bf k},t)=\frac 1{k_0}\pi _\mu({\bf k},t)+2\omega
_{[\mu \nu ]}q_\nu({\bf k},t)+
\]
\begin{equation}
+\frac 2{k_0}\left[ \omega _{(\mu \nu )}-\frac 14\omega _{(\alpha \alpha
)}\delta _{\mu \nu }\right] \pi _\nu ({\bf k},t),  \label{22}
\end{equation}
\[
\delta \pi _\mu({\bf k},t)=-k_0\omega _0q_\mu ({\bf k},t)+2\omega
_{[\mu \nu ]}\pi _\nu({\bf k},t)-
\]
\begin{equation}
-2k_0\left[ \omega _{(\mu \nu )}-\frac 14\omega _{(\alpha \alpha )}\delta
_{\mu \nu }\right] q_\nu ({\bf k},t).  \label{23}
\end{equation}

To get integrals of motion we use the method of generating functions [14].
It is verified that the generating function corresponding to transformations
(22), (23) is
\[
F=\sum_{{\bf k}}\biggl \{q_\mu ({\bf k},t)\pi _\mu ^{\prime }(
{\bf k},t)+\frac{\omega _0}2\left[ \frac{\left( \pi _\mu ^{\prime }(
{\bf k},t)\right) ^2}{k_0}+k_0\left( q_\mu ({\bf k},t)\right)
^2\right] +
\]
\[
+\omega _{[\mu \nu ]}\left[ \pi _\mu ^{\prime }({\bf k},t)q_\nu (
{\bf k},t)-\pi _\nu ^{\prime }({\bf k},t)q_\mu ({\bf k}
,t)\right] +
\]
\begin{equation}
+\left( \omega _{(\mu \nu )}-\frac 14\omega _{(\alpha \alpha )}\delta _{\mu
\nu }\right) \left[ \frac{\pi _\mu ^{\prime }({\bf k},t)\pi _\nu
^{\prime }({\bf k},t)}{k_0}+k_0q_\mu ({\bf k},t)q_\nu ({\bf k}
,t)\right] \biggr \},  \label{24}
\end{equation}

so that $q_\mu ^{\prime }({\bf k},t)=\partial F/\partial \pi _\mu
^{\prime }({\bf k},t)$, $\pi _\mu ({\bf k},t)=\partial F/\partial
q_\mu ({\bf k},t)$. From Eq.(24) we find the conservation law of the
following tensors
\[
J_{[\mu \nu ]}=\sum_{{\bf k}}\left[ \pi _\mu ({\bf k},t)q_\nu (
{\bf k},t)-\pi _\nu ({\bf k},t)q_\mu ({\bf k},t)\right] =
\]
\begin{equation}
=i\sum_{{\bf k}}\left[ b_\mu ^{+}({\bf k})b_\nu ({\bf k})-b_\nu
^{+}({\bf k})b_\mu ({\bf k})\right] ,  \label{25}
\end{equation}
\[
J_{(\mu \nu )}=\sum_{{\bf k}}\biggl \{\frac{\pi _\mu ({\bf k},t)\pi
_\nu ({\bf k},t)}{k_0}+k_0 q_\mu ({\bf k},t)q_\nu ({\bf k},t)-
\]
\[
-\frac {1}{4}\delta _{\mu \nu }\left[ \frac{\left( \pi _\alpha ({\bf k}
,t)\right) ^2}{k_0}+k_0\left( q_\alpha ({\bf k},t)\right) ^2\right]
\biggr \}=
\]
\begin{equation}
=\sum_{{\bf k}}\left[ b_\mu ^{+}({\bf k})b_\nu ({\bf k})+b_\nu ^{+}(
{\bf k})b_\mu ({\bf k})-\frac 12 \delta _{\mu \nu }\left( b_\alpha ^{+}(
{\bf k})b_\alpha ({\bf k})\right) \right] ,  \label{26}
\end{equation}
\begin{equation}
J=\frac 12 \sum_{{\bf k}}\left[ \frac{\left( \pi _\alpha ({\bf k}
,t)\right) ^2}{k_0}+k_0\left( q_\alpha ({\bf k},t)\right) ^2\right]
=\sum_{{\bf k}}b_\mu ^{+}({\bf k})b_\nu ({\bf k}),  \label{27}
\end{equation}

where variables $b_\mu ({\bf k})=\sqrt{2k_0}B_\mu ({\bf k})$ and
$b_\mu ^{+}({\bf k})=\sqrt{2k_0}B_\mu ^{+}({\bf k})$ in the quantized
theory are annihilation and creation operators, respectively. The conserved
variables (25)-(27) satisfy the equations $\left\{ J_{_{[\mu \nu
]}},H\right\} =\left\{ J_{_{(\mu \nu )}},H\right\} =\left\{ J,H\right\} =0$,
where $H$ is given by (14) and the $\left\{ ...\right\}$ are the
classical Poisson brackets. In the quantum case, operators $b_\mu ({\bf k}
)$, $b_\nu ^{+}({\bf k})$ obey the commutation relation $\left[ b_\mu (
{\bf k}),b_\nu ^{+}({\bf k}^{\prime })\right] =\delta _{\mu \nu
}\delta \left( {\bf k}-{\bf k}^{\prime }\right)$ and the $J_{_{[\mu
\nu ]}}$, $J_{_{(\mu \nu )}}$, $J$ are the generators of the group of
internal symmetry $U(3,1)$ so that $\left[ J_{_{[\mu \nu ]}},H\right]
=\left[ J_{_{(\mu \nu )}},H\right] =\left[ J,H\right] =0$. Using the
expansion
\[
B_\mu (x)=L^{-3/2}\sum_{{\bf k}}\left[ B_\mu ({\bf k}
)e^{i\left({\bf kx}-k_0t\right) }+B_\mu ^{+}({\bf k})e^{-i\left(
{\bf kx}-k_0 t\right)}\right] ,
\]

it is easy to verify that the integral of motion, $J_{_{[\mu \nu ]}}$
(Eq.(25)), coincides with the conserved spin momentum tensor:
\begin{equation}
J_{_{[\mu \nu ]}}=i\int S_{4[\mu \nu ]}d^3x.  \label{28}
\end{equation}

So the group $SO(3,1)$ under consideration ($\Lambda $) with the generators
$J_{_{[\mu \nu ]}}$ is the subgroup of the general group $U(3,1)$. The
conserved quantity $J$ is the number of quanta of the field. The generator
$J$ defines the subgroup $U(1)$ of the phase transformations and the
generators $J_{_{[\mu \nu ]}}$, $J_{_{(\mu \nu )}}$ correspond to the group
$SU(3,1)$. In the general case, transformations (16) in the coordinate
space are integro-differential transformations and therefore the Lagrangian
formalism is not convenient for studying this symmetry. The use of the
canonical formalism and the method of generating functions allow us to
investigate in a simple manner the group of ``addition'' symmetry of the
vector field which possesses two spin values $0,1$ (massive and massless
fields). The analogous group of
symmetry $SU(n)$ of the $n$-dimensional oscillator has been investigated
[11, 12]. The analysis performed is readily generalized to arbitrary
vector fields.

So the integral of motion (25)-(27) corresponds to the
transformations of internal-symmetry (16) which are not induced by
the space-time transformations.

The canonical variables $q_\mu (x)$, $\pi _\mu (x)$ satisfy the commutation
relation
\begin{equation}
\left\{ q_\mu (x),\pi _\nu (x^{\prime })\right\} _{t=t^{\prime }}=\delta
_{\mu \nu }\delta \left( {\bf x}-{\bf x}^{\prime }\right) .
\label{29}
\end{equation}

Using Eq.(7) we arrive at the relationships
\begin{equation}
\left\{ q_\mu (x),\stackrel{\cdot }{q}_\nu (x^{\prime })\right\}
_{t=t^{\prime }}=\delta _{\mu \nu }\delta \left( {\bf x}-{\bf x}
^{\prime }\right) .  \label{30}
\end{equation}

To quantize the fields we
should transfer to quantum commutators in accordance with the procedure
$\left\{ .,.\right\} _{t=t^{\prime }}\rightarrow i[.,.]$, where $[q,\pi
]=q\pi -\pi q$. Taking into account Eqs.(6), (30) we find the
commutation relations
\begin{equation}
\left[ B_\mu (x),\stackrel{\cdot }{B}_\nu (x^{\prime })\right] _{t=t^{\prime
}}=i\delta _{\mu \nu }\delta \left( {\bf x}-{\bf x}^{\prime }\right) ,
\label{31}
\end{equation}

It is obvious that commutators (31) correspond to Bose-Einstein statistics.

\section{Quantized fields and indefinite metric}

Now we will consider the quantized theory of the fields with multi-spin 0,1.
In accordance with the general rule that the
translation generator $P_\mu $ and the generator of four-dimensional
rotations $M_{\rho \sigma }$ are given by [9, 10]
\[
P_\mu =i\int T_{4\mu }d^3x,
\]
\begin{equation}
M_{\rho \sigma }=-i\int M_{4[\rho \sigma ]}d^3x.  \label{32}
\end{equation}

Using Eqs.(5) and commutation relations (31), it is not
difficult to check that generators (32) satisfy the commutation
relations of the Poincar\'e group.

Let us discuss the eigenvalues of the operator of the energy
$P_0=-iP_{4}$ of the fields considered. In the momentum representation, the
energy operator has the form:
\begin{equation}
P_0=\int k_0 \left( b_a ^{+}({\bf k})b_a ({\bf k})-
b_0 ^{+}({\bf k})b_0 ({\bf k})\right) d^3k,  \label{33}
\end{equation}

where operators
\[
b_\mu ^{+}({\bf k})=\sqrt{2k_0}B_\mu ^{+}({\bf k}),\hspace{0.2in}b_\mu
({\bf k})=\sqrt{2k_0}B_\mu ({\bf k})\hspace{0.2in}\left( B_\mu (
{\bf k},t)=B_\mu ({\bf k})\exp (-ik_0t)\right) ,
\]
as follows from Eqs.(31), must obey the following commutation relations:
\begin{equation}
\left[ b_\mu ({\bf k}),b_\nu ^{+}({\bf k}^{\prime })\right] =\delta
_{\mu \nu }\delta \left({\bf k}-{\bf k}^{\prime }\right).
\label{34}
\end{equation}
Commutation relation (34) is invariant under the group of
symmetry (16) because the operator of finite transformations $U$
is unitary. It is seen from Eq.(33) that the field energy, in
classical theory, is not positive-definite. In commutation relations (34)
there is a minus sign (as $b_4({\bf k})=ib_0({\bf k})$,
$b_4^{+}({\bf k})=ib_0^{+}({\bf k})$), and the operators
$b_0({\bf k})$, $b_0^{+}({\bf k})$ obey
the ``incorrect'' commutation relation
\[
\left[ b_0({\bf k}),b_0^{+}({\bf k}^{\prime })\right] =-\delta \left(
{\bf k}-{\bf k}^{\prime }\right) .
\]

From this equation it is seen that the term which appears in the
energy operator with a ($-$) sign satisfies the ``incorrect'' commutation
relation. The ``incorrect'' commutation
relations are not compatible, however,
with the assumption that the fields are real. As in the case of the
electromagnetic field, this difficulty is surmounted by introducing an
indefinite metric (see e.g., [15]). There are two possibilities. According
to the first one we can consider operators $b_0({\bf k}),$ $b_a^{+}(
{\bf k}^{\prime }),$ as the
creation operators, and $b_a({\bf k}),$ $b_0^{+}({\bf k}^{\prime
})$, as the annihilation operators of
particles. The vacuum state $\mid 0\rangle _1$ is defined by the
requirement:
\[
b_a({\bf k})\mid 0\rangle _1=b_0^{+}({\bf k})\mid 0\rangle _1=0,
\]
\begin{equation}
_1\langle 0\mid 0\rangle _1=1.  \label{35}
\end{equation}

The basis in Hilbert space is given by
\begin{equation}
\mid m,n\rangle _1=\frac 1{\sqrt{m!n!}}\left( b_{a}^{+}({\bf k})\right)
^m\left( b_{0}({\bf k})\right) ^n\mid 0\rangle _1  \label{36}
\end{equation}

with the normalization condition
\begin{equation}
\langle m,n\mid m^{\prime },n^{\prime }\rangle _1=\delta _{mm^{\prime
}}\delta _{nn^{\prime }}.  \label{37}
\end{equation}

In this case there is Hilbert space but the eigenstates of the energy operator
(33) are not positive defined and the interpretation of the states with
negative energy is problematical.

The second possibility [15] is more favorable and connected with
introducing an indefinite metric. In this case the algebra (34) can
be represented as the operator algebra in the states with the indefinite
metric and here operators $b_\mu ^{+}({\bf k})$ are the creation and
$b_\mu ({\bf k})$ the annihilation operators. As the basis of the irreducible
representation in Hilbert space we choose the following vectors:
\[
\mid m,n\rangle =\frac 1{\sqrt{m!n!}}\left( b_{a}^{+}({\bf k})\right)^m
\left( b_{0}^{+}({\bf k})\right)^n
\mid 0\rangle ,
\]
\begin{equation}
b_{\mu}({\bf k})\mid 0\rangle =0,\hspace{0.5in}
\langle 0\mid 0\rangle =1.  \label{38}
\end{equation}

With the help of Eqs.(34) we arrive at
\begin{equation}
\langle m,n\mid m^{\prime },n^{\prime }\rangle =\left( -1\right) ^n\delta
_{mm^{\prime }}\delta _{nn^{\prime }}.  \label{39}
\end{equation}

In this way, the state space of vector fields has an indefinite metric, i.e.
the square vector norm can be negative. Using Eqs.(34), it is easy
to verify that the eigenvalues of the operator $P_0$ are positive; however,
the metric is still indefinite.

As usual, in the theory with an indefinite metric, it is necessary to divide
the space into ``physical'' and ``nonphysical'' subspaces (see [15, 16]).
The ``physical'' subspace corresponds to a positive square norm, and the
``nonphysical'' to a negative square norm. The state vectors $\mid
m,0\rangle $ create the ``physical'' subspace $H_p$ and $\mid m,n\rangle$
(at $n\neq 0$) create the ``nonphysical'' subspace $H_n$. The physical
states permit the usual probability interpretation. It is apparent from
Eq.(39) that the ``physical'' and ``nonphysical'' subspaces are orthogonal
$\langle m,0\mid m^{\prime },n^{\prime }\rangle =0$, and therefore the state
space is the direct sum of the two subspaces $H_p$ and $H_n$. We can then
represent any vector in the form
\begin{equation}
\mid \rangle =\mid \rangle _p+\mid \rangle _n,  \label{40}
\end{equation}

where $\mid \rangle _p\in H_p$, $\mid \rangle _n\in H_n$. In scattering
processes, states $\mid in\rangle $, $\mid out\rangle $ should belong to the
``physical'' subspace $H_p$. It is possible also to consider transitions
between states which belong to the ``nonphysical'' subspace $H_n$. But the
consideration of transitions between the vectors $\mid \rangle _p$, $\mid
\rangle _n$ presents difficulties [15] because such transitions violate the
unitarity of the $S$-matrix and the usual probability interpretation.

\section{Internal symmetry of the electromagnetic field}

It will be shown that the symmetry-group of an electromagnetic field is
$U(2)$.
When we impose the constraints $\partial_\mu B_\mu=0$ (The Lorentz condition),
$m=0$, we arrive at the transition to the formulation of electrodynamics,
and the Hamiltonian (33) takes the form
\begin{equation}
H=\sum_{{\bf k}}k_0\left( b_1^{+}({\bf k})b_1({\bf k})+b_2^{+}(
{\bf k})b_2({\bf k})\right) .  \label{41}
\end{equation}

In this case the Hamiltonian (41) is invariant under the transformations
of the $U(2)$ group:
\begin{equation}
\left(
\begin{array}{c}
b_1^{\prime }({\bf k}) \\
b_2^{\prime }({\bf k})
\end{array}
\right) =\exp \left( \frac i2\alpha +i{\bf n\tau }\frac \theta 2\right)
\left(
\begin{array}{c}
b_1({\bf k}) \\
b_2({\bf k})
\end{array}
\right) ,  \label{42}
\end{equation}

where ${\bf n}^2=1$, $\alpha $, $\theta $ are real group parameters,
${\bf \tau}$ are the Pauli matrices. So,
the internal symmetry group $U(3,1)$ being considered, after the transition to
electrodynamics, reduces to the $U(2)$ group. Using the procedure described
above, after finding the generalizing function, we come to the following
conserved quantities:
\[
J_1=\frac 12\sum_{{\bf k}}\left[ b_1^{+}({\bf k})b_2({\bf k}
)+b_2^{+}({\bf k})b_1({\bf k})\right] ,
\]
\[
J_2=\frac i2\sum_{{\bf k}}\left[ b_2^{+}({\bf k})b_1({\bf k}
)-b_1^{+}({\bf k})b_2({\bf k})\right] ,
\]
\begin{equation}
J_3=\frac 12\sum_{{\bf k}}\left[ b_1^{+}({\bf k})b_1({\bf k}
)-b_2^{+}({\bf k})b_2({\bf k})\right] ,  \label{43}
\end{equation}
\begin{equation}
J_0=\frac 12\sum_{{\bf k}}\left[ b_1^{+}({\bf k})b_1({\bf k})+b_2^{+}(
{\bf k})b_2({\bf k})\right] .  \label{44}
\end{equation}

The generators $J_i$ satisfy the commutation relation of the $SU(2)$ group,
and commute with $J_0$: $\left[ J_i,J_0\right] =0$. There is a representation of
the rotation generators in the Hilbert space in the form of Eqs.(43) in
[17, 18]. The average values of the operators (43) are identified in
[19] with the Stokes parameters characterizing the polarization of the
electromagnetic wave.

It is obvious, that the transformations (42), and the integrals of
motion (43), (44), can be written in coordinate space but in
nonlocal form. So, in the coordinate representation the expression
(44) takes the nonlocal form [20]
\begin{equation}
J_0=\frac 1{2(2\pi )^3}\int \int \frac{{\bf E}{\bf (x)E}{\bf (y)
}+{\bf H}{\bf (x)H}({\bf y})}{\mid {\bf x}-{\bf y}\mid ^2}
d^3xd^3y.  \label{45}
\end{equation}

The transformations (42) at $\theta =0$, which lead to the integral
of motion (45) are local only in the momentum representation, and in the
coordinate representation are integro-differential transformations.

It should be noted that the transformations (42) with $
n_1=n_3=0$, $\alpha =0$, $n_2=1,$ lead to the dual transformations of the
strengths of electric ${\bf E,}$ and magnetic ${\bf H}$ fields:
\[
{\bf E}^{\prime }={\bf E}\cos \frac \theta 2+{\bf H}\sin \frac
\theta 2,
\]
\begin{equation}
{\bf H}^{\prime }={\bf H}\cos \frac \theta 2-{\bf E}\sin \frac
\theta 2.  \label{46}
\end{equation}

Hence, the well known group of dual transformations is the subgroup of the $
SU(2)$ group under consideration.

\section{ First-order equations and density matrix}

Now we consider the matrix formulation of the first-order of the
Stueckelberg fields (3) which is convenient for constructing the density matrix
and for some electrodynamics calculations. All the linearly independent
solutions of the equation for a free particle will be obtained in terms of
the projection matrix-dyads.

Let us introduce the 11-dimensional function
\begin{equation}
\Psi (x)=\left\{ \psi _A(x)\right\} =\frac 1m\left(
\begin{array}{c}
-\psi _0 \\
m\psi _\mu \\
\psi _{[\mu \nu ]}
\end{array}
\right) \hspace{0.5in}(A=0,\mu,[\mu \nu ]),  \label{47}
\end{equation}

where $\mu ,$ $\nu =1,$ $2,$ $3,$ $4$. Using the elements of the entire
algebra Eq. (19), equations (3) can be written in the form of one
equation
\[
\partial _\nu \left( \varepsilon ^{\mu ,[\mu \nu ]}+\varepsilon ^{[\mu \nu
],\mu }+\varepsilon ^{\nu ,0}+\varepsilon ^{0,\nu }\right) _{AB}\Psi _B(x)+
\]
\begin{equation}
+m\left[ \varepsilon ^{\mu ,\mu }+ \varepsilon ^{0,0}+\frac
12\varepsilon ^{[\mu \nu ],[\mu \nu ]}\right] _{AB}\Psi _B(x)=0.
\label{48}
\end{equation}

After introducing 11-dimensional matrices
\[
\alpha _\nu =\varepsilon ^{\mu ,[\mu \nu ]}+\varepsilon ^{[\mu \nu ],\mu
}+\varepsilon ^{\nu ,0}+\varepsilon ^{0,\nu },
\]
\begin{equation}
I_{11}=\varepsilon ^{\mu ,\mu }+\varepsilon
^{0,0}+\frac 12\varepsilon ^{[\mu \nu ],[\mu \nu ]},  \label{49}
\end{equation}

Eq. (48) takes the form of the relativistic wave equation of the first
order:
\begin{equation}
\left( \alpha _\mu \partial _\mu +m\right) \Psi (x)=0.
\label{50}
\end{equation}
We took into account that $I_{11}$ in Eq. (49) is the unit matrix in $11-$dimensional space.

Eq. (50) represents the Stueckelberg equation for massive fields in the
matrix form. When fields $\Psi_A (x)$ are complex values, Eq. (50) describes charged particles
with multi-spin $0, 1$.

It should be noted that the matrices $\alpha _\mu $ can be
represented as
\[
\alpha _\mu =\beta _\mu ^{(1)}+\beta _\mu ^{(0)},
\]
\[
\beta _\nu ^{(1)}=\varepsilon ^{\mu ,[\mu \nu ]}+\varepsilon ^{[\mu \nu
],\mu },
\]
\begin{equation}
\beta _\nu ^{(0)}=\varepsilon ^{\nu ,0}+\varepsilon ^{0,\nu },  \label{51}
\end{equation}

where the $10-$dimensional $\beta _\mu ^{(1)}$ and $5-$dimensional $\beta
_\mu ^{(0)}$ matrices obey the Petiau-Duffin-Kemmer [21-23] algebra:
\begin{equation}
\beta _\mu \beta _\nu \beta _\alpha +\beta _\alpha \beta _\nu \beta _\mu
=\delta _{\mu \nu }\beta _\alpha +\delta _{\alpha \nu }\beta _\mu ,
\label{52}
\end{equation}

so that the equations for spin-$1$ and spin-$0$ particles are
\begin{equation}
\left( \beta _\mu ^{(1)}\partial _\mu +m\right) \Psi ^{(1)}(x)=0,
\hspace{0.5in}\Psi ^{(1)}(x)=\frac 1m \left(
\begin{array}{c}
m\psi _\mu \\
\psi _{[\mu \nu ]}
\end{array}
\right) ,  \label{53}
\end{equation}
\begin{equation}
\left( \beta _\mu ^{(0)}\partial _\mu +m\right) \Psi ^{(0)}(x)=0,
\hspace{0.5in}\Psi ^{(0)}(x)=\frac 1m \left(
\begin{array}{c}
-\psi _0 \\
m\psi _\mu
\end{array}
\right) .  \label{54}
\end{equation}

The $10-$dimensional Petiau-Duffin-Kemmer equation (53) is equivalent to
the Proca equations [24] for spin-$1$ particles and the $5-$dimensional Eq.
(54) is equivalent to the Klein-Gordon-Fock equation for scalar particles.
The $11-$dimensional Eq. (50) describes fields with two spins $0,$ $1$.
It is not difficult to verify
(using Eqs. (19)) that the $11-$dimensional matrices $\alpha _\mu $ (49)
satisfy the algebra (see also [25]):
\[
\alpha _\mu \alpha _\nu \alpha _\alpha +\alpha _\alpha \alpha _\nu \alpha
_\mu +\alpha _\mu \alpha _\alpha \alpha _\nu +\alpha _\nu \alpha _\alpha
\alpha _\mu +\alpha _\nu \alpha _\mu \alpha _\alpha +\alpha _\alpha \alpha
_\mu \alpha _\nu =
\]
\begin{equation}
=2\left( \delta _{\mu \nu }\alpha _\alpha +\delta _{\alpha \nu }\alpha _\mu
+\delta _{\mu \alpha }\alpha _\nu \right) .  \label{55}
\end{equation}

This algebra is more complicated than the Petiau-Duffin-Kemmer algebra
(52). Different representations of the Petiau-Duffin-Kemmer algebra (52)
were considered in [26-28].

Now we find the solutions to Eq. (50) corresponding to definite values of
the energy and momentum of a quantum of the massive fields. In the momentum
space Eq. (50) becomes
\begin{equation}
-i\widehat{p}\Psi _p=\varepsilon m\Psi _p,  \label{56}
\end{equation}

where $\widehat{p}=\alpha _\mu p_\mu $, $p_\mu =({\bf p},ip_0)$, $p^2=
{\bf p}^2-p_0^2=-m^2$; the value of $\varepsilon =1$ corresponds to
positive energy and $\varepsilon =-1$ to negative energy. Here ${\bf p}$
means the momentum of a field-quantum. It may be verified using (55) that
the equality
\begin{equation}
\widehat{p}^3=p^2\widehat{p}  \label{57}
\end{equation}

is valid. Following the general method of projection operators [29, 30],
we find solutions to Eq. (56) in the form of the projection matrix
\begin{equation}
M_\varepsilon =\frac{i\widehat{p}\left( i\widehat{p}-\varepsilon m\right) }{
2m^2},  \label{58}
\end{equation}

so that
\begin{equation}
M_\varepsilon ^2=M_\varepsilon ,  \label{59}
\end{equation}
and $\varepsilon =\pm 1$. Every column of the matrix $M_\varepsilon $ can be
considered as an eigenvector $\Psi _p$ of equation (56) with eigenvalue $
\varepsilon m$. Eq. (59) for projection operators tells that matrix $
M_\varepsilon $ can be transformed into diagonal form, with the diagonal
containing only ones and zeroes. So the $M_\varepsilon $ acting on the wave
function $\Psi $ will retain components which correspond to the eigenvalue $
\varepsilon m$. The generators of the Lorentz group in the $11-$dimensional
space being considered are given by
\begin{equation}
J_{\mu \nu }=\beta _\mu ^{(1)}\beta _\nu ^{(1)}-\beta _\nu ^{(1)}\beta _\mu
^{(1)}.  \label{60}
\end{equation}

It should be noted that matrices (60) act in the $10-$dimensional
subspace $\left( m\psi _\mu ,\psi _{[\mu \nu ]}\right) $ because the scalar $
\psi _0$ is an invariant of the Lorentz transformations. So matrices (60)
are also generators of the Lorentz group for the Petiau-Duffin-Kemmer fields
of Eq. (53). Using properties (19), we get the commutation relations
\begin{equation}
\left[ J_{\rho \sigma },J_{\mu \nu }\right] =\delta _{\sigma \mu }J_{\rho
\nu }+\delta _{\rho \nu }J_{\sigma \mu }-\delta _{\rho \mu }J_{\sigma \nu
}-\delta _{\sigma \nu }J_{\rho \mu },  \label{61}
\end{equation}
\begin{equation}
\left[ \alpha _\lambda ,J_{\mu \nu }\right] =\delta _{\lambda \mu }\alpha
_\nu -\delta _{\lambda \nu }\alpha _\mu .  \label{62}
\end{equation}

Relationship (61) is a well known commutation relation for generators of
the Lorentz group $SO(3,1)$. Equation (50) is form-invariant under the
Lorentz transformations since relation (62) is valid. To guarantee the
existence of a relativistically invariant bilinear form
\begin{equation}
\overline{\Psi }\Psi =\Psi ^{+}\eta \Psi ,  \label{63}
\end{equation}

where $\Psi ^{+}$ is the Hermitian-conjugate wave function, we should
construct a Hermitianizing matrix $\eta $ with the properties [9, 28,
30]:
\begin{equation}
\eta \alpha _i=-\alpha _i\eta ,\hspace{0.5in}\eta \alpha _4=\alpha _4\eta
\hspace{0.5in}(i=1,2,3).  \label{64}
\end{equation}

Such a matrix exists and is given by
\[
\eta =-\varepsilon ^{0,0}+2\beta _4^{(1)2}-I_{10},
\]
\begin{equation}
I_{10}=\varepsilon ^{\mu ,\mu }+\frac 12\varepsilon ^{[\mu \nu ],[\mu \nu ]},
\label{65}
\end{equation}

where the matrix $\eta ^{(1)}=2\beta _4^{(1)2}-I_{10}$ plays the role of a
Hermitianizing matrix for the Petiau-Duffin-Kemmer equation (53) [9].
The operator of the squared spin (squared Pauli-Lubanski vector) is given by
\begin{equation}
\sigma ^2=\left( \frac 1{2m}\varepsilon _{\mu \nu \alpha \beta }p_\nu
J_{\alpha \beta }\right) ^2=\frac 1{m^2}\left( J_{\mu \nu }^2p^2-J_{\mu
\sigma }J_{\nu \sigma }p_\mu p_\nu \right) .  \label{66}
\end{equation}

It may be verified that this operator obeys the minimal equation
\begin{equation}
\sigma ^2\left( \sigma ^2-2\right) =0,  \label{67}
\end{equation}

so that eigenvalues of the squared spin operator $\sigma ^2$ are $s(s+1)=0$
and $s(s+1)=2$. This confirms that the considered fields describe the
superposition of two spins $s=0$ and $s=1$. To separate these states we use
the projection operators
\begin{equation}
S_{(0)}^2=1-\frac{\sigma ^2}2,\hspace{0.5in}S_{(1)}^2=\frac{\sigma ^2}2
\label{68}
\end{equation}

having the properties $S_{(0)}^2S_{(1)}^2=0$, $\left( S_{(0)}^2\right)
^2=S_{(0)}^2$, $\left( S_{(1)}^2\right) ^2=S_{(1)}^2$, $
S_{(0)}^2+S_{(1)}^2=1 $, where $1\equiv I_{11}$ is the unit matrix in $11-$
dimensional space. In accordance with the general properties of the
projection operators, the matrices $S_{(0)}^2$, $S_{(1)}^2$ acting on the
wave function extract pure states with spin $0$ and $1$, respectively. Now
we introduce the operator of the spin projection on the direction of the
momentum $\mathbf{p}$ :
\begin{equation}
\sigma _p=-\frac i{2\mid \mathbf{p}\mid }\epsilon _{abc}p_aJ_{bc}=-\frac
i{\mid \mathbf{p}\mid }\epsilon _{abc}p_a\beta _b^{(1)}\beta _c^{(1)},
\label{69}
\end{equation}

where $\mid {\bf p}\mid =\sqrt{{\bf p}_1^2+{\bf p}_2^2+{\bf p}
_3^2}.$ The minimal matrix equation for the spin projection operator is
\begin{equation}
\sigma _p\left( \sigma _p-1\right) \left( \sigma _p+1\right) =0
\label{70}
\end{equation}

and the corresponding projection operators are given by
\begin{equation}
\widehat{S}_{(\pm 1)}=\frac 12\sigma _p\left( \sigma _p\pm 1\right)
\hspace{0.5in}\widehat{S}_{(0)}=1-\sigma _p^2.  \label{71}
\end{equation}

Operators $S_{(\pm 1)}$ correspond to the spin projections $s_p=\pm 1$ and $
S_{(0)}$ to $s_p=0$. It is easy to verify that the required commutation
relations hold:
\[
\left[ S_{(0)}^2,\widehat{p}\right] =\left[ S_{(1)}^2,\widehat{p}\right]
=\left[ \widehat{S}_{(\pm 1)},\widehat{p}\right] =\left[ \widehat{S}_{(0)},
\widehat{p}\right] =0,
\]
\begin{equation}
\left[ S_{(0)}^2,\widehat{S}_{(\pm 1)}\right] =\left[ S_{(1)}^2,\widehat{S}
_{(\pm 1)}\right] =\left[ S_{(0)}^2,\widehat{S}_{(0)}\right] =0.
\label{72}
\end{equation}

Thus the projection matrices extracting pure states with definite spin, spin
projection and energy take the form
\[
\Delta _{\varepsilon ,\pm 1}=M_\varepsilon S_{(1)}^2\widehat{S}_{(\pm 1)}=
\frac{i\widehat{p}\left( i\widehat{p}-\varepsilon m\right) }{2m^2}\frac
12\sigma _p\left( \sigma _p\pm 1\right) ,
\]
\[
\Delta _\varepsilon ^{(1)}=M_\varepsilon S_{(1)}^2\widehat{S}_{(0)}=\frac{i
\widehat{p}\left( i\widehat{p}-\varepsilon m\right) }{2m^2}\frac{\sigma ^2}
2\left( 1-\sigma _p^2\right) ,
\]
\begin{equation}
\Delta _\varepsilon ^{(0)}=M_\varepsilon S_{(0)}^2\widehat{S}_{(0)}=\frac{i
\widehat{p}\left( i\widehat{p}-\varepsilon m\right) }{2m^2}\left( 1-\frac{
\sigma ^2}2\right) \left( 1-\sigma _p^2\right) ,  \label{73}
\end{equation}

where we took into account that $\left( \sigma ^2/2\right) \sigma _p=\sigma
_p$. Projection operators $\Delta _{\varepsilon ,\pm 1}$, $\Delta
_\varepsilon ^{(1)}$ extract states with spin $1$ and spin projections $\pm
1 $, $0$, and $\Delta _\varepsilon ^{(0)}$ corresponds to spin $0$. The $
\Delta _{\varepsilon ,\pm 1}$, $\Delta _\varepsilon ^{(1)}$, $\Delta
_\varepsilon ^{(0)}$ are the density matrices for pure spin spates. It is
easy to consider impure states by summation of Eqs. (73) over spin
projections and spins. Projection operators for pure states can be
represented as matrix-dyads [29, 30]:
\begin{equation}
\Delta _{\varepsilon ,\pm 1}=\Psi _{\varepsilon ,\pm 1}\cdot \overline{\Psi }
_{\varepsilon ,\pm 1},\hspace{0.3in}\Delta _\varepsilon ^{(1)}=\Psi
_\varepsilon \cdot \overline{\Psi }_\varepsilon ,\hspace{0.3in}\Delta
_\varepsilon ^{(0)}=\Psi _{\varepsilon ,}^{(0)}\cdot \overline{\Psi }
_{\varepsilon ,}^{(0)},  \label{74}
\end{equation}

where the wave functions $\Psi _{\varepsilon ,\pm 1}$, $\Psi _\varepsilon $
are the solution of the field equations for spin $1$ and spin projections $
\pm 1$ and $0$, respectively, and $\Psi _\varepsilon ^{(0)}$ corresponds to
the solution with spin $0$. Expressions (73), (74) are convenient for
calculating different electrodynamics processes involving polarized vector charged
particles.

\end{document}